\documentclass[12pt]{article}
\usepackage{geometry}
\usepackage{amsmath,amssymb}
\usepackage[nosort]{cite}
\usepackage{float}
\usepackage{latexsym}
\usepackage{graphicx}
\usepackage{color}
\usepackage{subfig}

\newcommand{\Tr}{{\rm Tr\,}}

\renewcommand{\d}{{\rm d}} 
\newcommand{\beq}{\begin{equation}}
\newcommand{\be}{\begin{equation}}
\newcommand{\ee}{\end{equation}}
\newcommand{\bea}{\begin{eqnarray}}
\newcommand{\eea}{\end{eqnarray}}

\newcommand{\pa}{\partial}
\newcommand{\pab}{\bar{\partial}}

\newcommand{\nn}{\nonumber}
\newcommand{\zb}{\bar{z}}

\newcommand{\Ab}{\bar{A}}

\newcommand{\Zb}{\bar{Z}}
\newcommand{\ga}{\gamma}
\newcommand{\sn}{{\rm sn}}
\newcommand{\cn}{{\rm cn}}
\newcommand{\dn}{{\rm dn}}
\newcommand{\mcH}{{\mathcal H}}

\begin{document}
\begin{titlepage}
\hbox to \hsize{\hspace*{0 cm}\hbox{\tt }\hss
    \hbox{\small{\tt }}}

\vspace{1 cm}

\centerline{\bf \Large Phase transitions in Wilson loop correlator  }

\vspace{.6cm}

\centerline{\bf \Large \\  from integrability in global AdS.}

\vspace{.6cm}


\vspace{1 cm}
 \centerline{\large  $^\dagger\!\!$ Benjamin A. Burrington and $^\star\!\!$ Leoplodo A. Pando Zayas}

\vspace{0.5cm}
\centerline{\it ${}^\dagger$Department of Physics,}
\centerline{\it University of Toronto,}
\centerline{\it Toronto, Ontario, Canada M5S 1A7. }

\vskip .5cm
\centerline{\it ${}^\star$ Michigan Center for Theoretical
Physics}
\centerline{ \it Randall Laboratory of Physics, The University of
Michigan}
\centerline{\it Ann Arbor, MI 48109-1040}

\begin{abstract}
We directly compute Wilson loop/Wilson loop correlators on ${\mathbb R}\times $S$^3$ in AdS/CFT by constructing space-like minimal surfaces that connect two space-like circular contours on the boundary of global AdS that are separated by a space-like interval.  We compare these minimal surfaces to the disconnected ``double cap'' solutions both to regulate the area, and show when the connected/disconnected solution is preferred.  We find that for sufficiently large Wilson loops no transition occurs because the Wilson loops cannot be sufficiently separated on the sphere.  This may be considered an effect similar to the Hawking-Page transition: the size of the sphere introduces a new scale into the problem, and so one can expect phase transitions to depend on this data.  To construct the minimal area solutions, we employ a reduction a la Arutyunov-Russo-Tseytlin (used by them for spinning strings), and rely on the  integrability of the reduced set of equations to write explicit results.
\end{abstract}
\end{titlepage}


\section{Introduction}


AdS/CFT \cite{Maldacena:1997re} has become an important tool for studying aspects of strongly coupled gauge theories (for a review, see \cite{Aharony:1999ti}).  In this context, classical minimal area worldsheets have played an interesting role: computing Wilson loop expectation values and quark/anti-quark potentials \cite{Rey:1998ik,Drukker:2000rr,Berenstein:1998ij};
computing the cusp anomalous dimension \cite{Gross:1998gk,Kruczenski:2002fb,Kruczenski:2007cy}; computing aspects of large spin or large R-charge operators \cite{classicalStringSpin,Frolov:2003qc,Arutyunov:2003za,Arutyunov:2003uj}; describing the behavior of string vertex operator correlators for an AdS target space \cite{Buchbinder:2010vw,Roiban:2010fe,Ryang:2010bn,Hernandez:2010tg}; and computing color ordered scattering data \cite{GluonScat} using the dual piecewise lightlike Wilson loop.
There has been great progress made in this area due to the integrability of the classical worldsheets on AdS (and sphere) backgrounds \cite{Arutyunov:2003za,Arutyunov:2003uj} as well as for the descendent Pohlmeyer reduced theories \cite{Pohlmeyer:1975nb}.

These computations often come down to evaluating a regularized worldsheet area and interpreting the exponentiation of this area as being some physical quantity of interest.  We first define the Wilson loop operator as a path ordered exponentiation of the gauge field along a closed contour $C$
\be
W[C]=\frac{1}{N}\Tr\left({\mathcal P} \exp\left(\int_C A_\mu dx^\mu\right)\right).
\ee
The full ${\mathcal N}=4$ super Yang-Mills (SYM) theory Wilson loop is discussed in \cite{Drukker:2000rr} and may contain some dependence on the scalar fields as well, however for our discussion we will will set this part to zero.  The expectation value of the Wilson loop operator is identified with the appropriately normalized exponentiated classical area of a world sheet ending on the contours $C$ at the boundary of AdS:
\be
\langle W[C]\rangle = \exp\left(-A[C]\right) \label{introVev}.
\ee
Since we have set the scalar part of the Wilson loop operator to zero, we will be studying world sheets that have a profile in the AdS only.  The solutions we will discuss are therefore applicable in any theory with an AdS factor in the geometry, however, one may need to be careful about the $g_s$ corrections to these backgrounds, as we shall discuss in a moment.

In what follows, we will be concerned with two Wilson loop operators, defined by two circular contours.  A natural object to study is \cite{Berenstein:1998ij,Zarembo:1999bu},
\be
\frac{\langle W[C_1] W[C_2] \rangle}{\langle W[C_1]\rangle \langle W[C_2] \rangle}=1+\sum_n\frac{\lambda^{2n}}{N^{2n}}f_n(\lambda,C_1,C_2). \label{wilsonratio}
\ee
These have been previously studied for Minkowski space holographically in \cite{Zarembo:1999bu,Olesen:2000ji,Drukker:2005cu,Ahn:2006px}, and see \cite{Plefka:2001bu}
 for a field theory discussion.  One can argue for the above expansion simply using large $N$ arguments.
The first two terms, $1$ and $\frac{1}{N^2}$, are in some sense planar terms, where the $1$ is enhanced due to being a ``disconnected'' diagram, and the order $\frac{1}{N^2}$ term comes from connected planar diagrams.  Higher order terms come from non-planar diagrams.  Further, we have extracted a certain power of $\lambda$ to make this look like a $g_s$ expansion, but is otherwise artificial, and could be absorbed into the definition of $f_n$.

\begin{figure}[!ht]
\begin{center}
\includegraphics[width=0.4\textwidth]{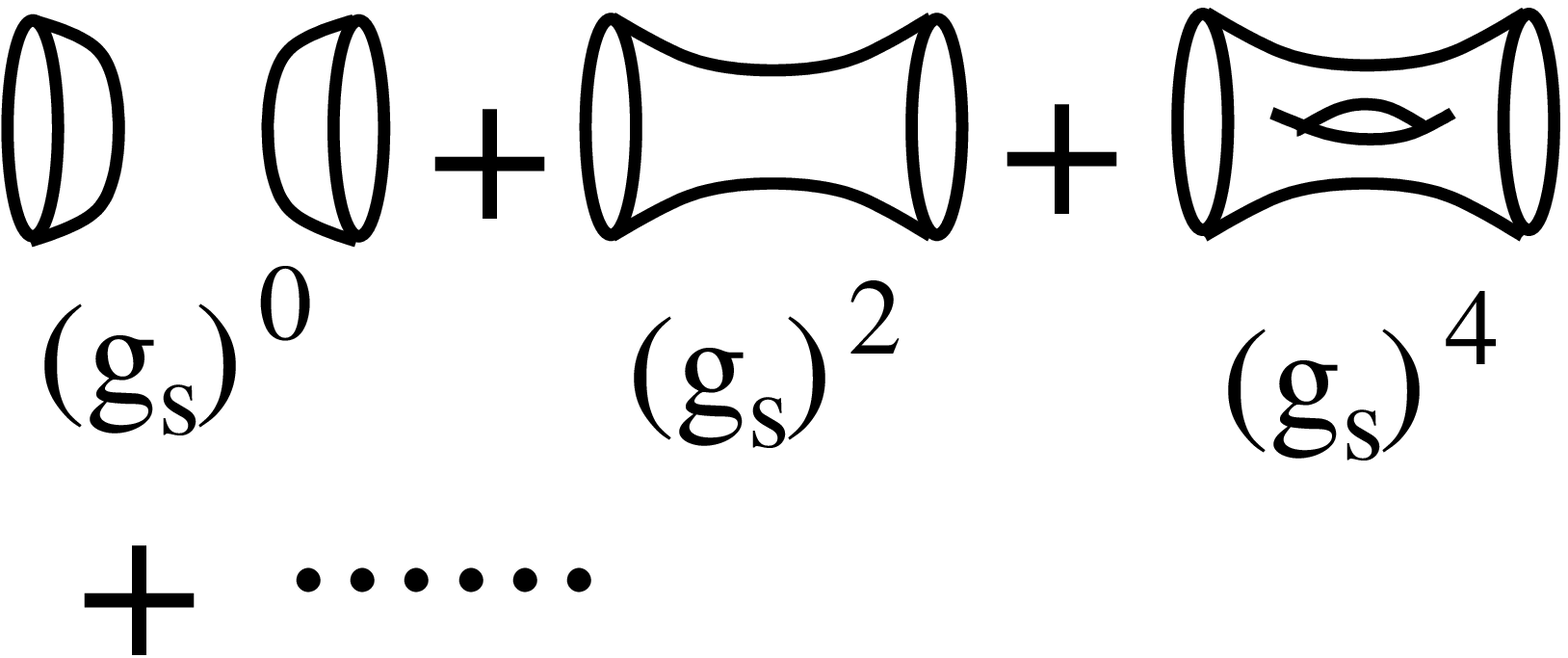}
\end{center}
\caption{A diagramatic expansion of the worldsheets living in AdS.  We have normalized the first term to be order one ($(g_s)^0$), so that the subsequent labels correspond to the degree to which they are sub-leading.  }
\label{diagExpansion}
\end{figure}

In the string picture, we expect to compute a set of diagrams given in figure \ref{diagExpansion}.
The fact that we want to compute a ratio of the form (\ref{wilsonratio}) means that we want to regulate all of the areas of the above worldsheets by subtracting off the area of the disconnected leading order piece, i.e. to compute
\be
\frac{\langle W[C_1] W[C_2] \rangle}{\langle W[C_1]\rangle \langle W[C_2] \rangle}=1+\sum_n g_s^{2n}\exp\left[-\left(A^n_{\rm connected}[C_1,C_2]-A^0_{\rm disconnected}[C_1,C_2]\right)\right]
\ee
where $A^n_{\rm connected}$ is the connected diagram with Euler character $-2n+2$.  The disconnected part is actually two distinct parts, coming from the area of the two disconnected pieces.  We recognize this expansion as being the same as the previous expansion because $g_s=g_{\rm YM}^2= \frac{\lambda}{N}$, and the exponentiated area plays the role of the function $f_n$.

Here we have relied on the S$^5$ in the AdS$_5\times$S$^5$ geometry in a subtle way: we have assumed that the background has no $g_s$ corrections.  Hence, for more general setups, like Sasaki-Einstein or orbifold backgrounds, more care may be needed; see \cite{Liu:2010gz} for recent work in this direction.  In such cases, there is a second type of $g_s$ correction coming from the corrected worldsheet action due to the $g_s$ correction to the background.  Therefore, lower genus diagrams will contribute to a given order in $g_s$.  For us, we will be discussing $f_1$, and so this correction would amount to evaluating the old double cap solution in the correction to the sigma model action.  In fact, because we divide by a pair of $\langle W \rangle$ expectation values, we expect that this correction is in fact divided out and so we expect that the above is sufficient for leading $\frac{1}{N^2}$ order corrections in any background with an AdS.  However, for higher corrections, one would need to be more careful, and check the contributions of lower genus worldsheets in the corrected action, as well as needing to modify the equations one solves.

\begin{figure}[!ht]
\centering
\includegraphics[width=0.3\textwidth]{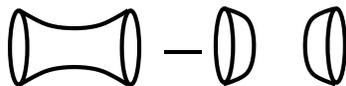}
\caption{ A diagramatic expansion of the first non trivial term we will compute.  This gives the function $f_1$ at strong coupling.}
\label{cyldiag}
\end{figure}

\begin{figure}[!ht]
\centering
\includegraphics[width=0.4\textwidth]{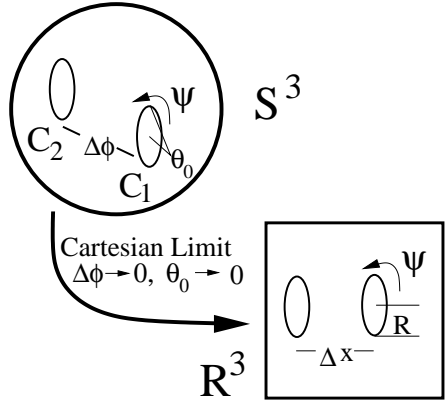}
\caption{Here we diagram the boundary conditions specifying the two equal size contours we use to explicitly calculate the Gross-Ooguri transition.  The Cartesian limit gives the flat space behavior.}
\label{bndycondfig}
\end{figure}

The purpose of this paper is to compute the first correction term in the above expansion in global AdS coordinates, namely, that associated with the ``diagram'' in figure \ref{cyldiag}.
We compute this analytically for two synchronous coaxial same-size spacelike Wilson loops, separated by an interval in one of the angular coordinates ($\phi$) of the S$^3$ inside of AdS$^5$, and we find
\bea
&&\ln(f_1)=-\left(A_{\rm connect}-A_{{\rm cap}\times 2}\right)\nn \\
&&\qquad \qquad \qquad \qquad =-\sqrt{4\pi\lambda}\left(1- \frac{1}{\sqrt{2k^2-1}}\left[-(1-k^2)K(k)+E(k)\right]\right)
\eea
with $k$ defined through the following relation
\bea
&& \exp\left({\frac{2k\sqrt{1-k^2}}{\sqrt{2k^2-1}}\left(K(k)-\frac{1-k^2}{k^2}\Pi\left(\frac{2k^2-1}{k^2},k\right)\right)}\right) \nn \\ &=& \frac{\sqrt{\sin^2(\theta_0)+\tan^2\left(\frac{\Delta\phi}{2}\right)}+\left|\tan\left(\frac{\Delta\phi}{2}\right)\right|\cos(\theta_0)} {\sqrt{\sin^2(\theta_0)+\tan^2\left(\frac{\Delta\phi}{2}\right)}-\left|\tan\left(\frac{\Delta\phi}{2}\right)\right|\cos(\theta_0)}  \label{introktobndydata} \eea
and where $\theta_0$ defines the ``radius'' of the two Wilson lines, and $\Delta \phi$ determines the distance between their center points (see fig. \ref{bndycondfig}).
The functions $K$ and $\Pi$ (and $E$ later on) are complete elliptic integrals, see appendix \ref{ellipticSection}.  The above is valid for $k<k_c$ where this critical value is given by
\be
\left(1- \frac{1}{\sqrt{2k_c^2-1}}\left[-(1-k_c^2)K(k_c)+E(k_c)\right]\right)=0
\ee
which is numerically found to be
\be
k_c=0.8232.
\ee
Using this value of $k_c$, one finds that the Gross-Ooguri transition happens at a $\Delta \phi$ defined by
\be
\tan^2\left(\frac{\Delta\phi_c}{2}\right)=\frac{\sin^2(\theta_0)}{\frac{({\mathcal E}+1)^2}{({\mathcal E}-1)^2}\cos^2(\theta_0)-1}
\ee
where ${\mathcal{E}}\approx 2.4034$ \footnote{${\mathcal E}$ is just the value of $\exp\left({\frac{2k\sqrt{1-k^2}}{\sqrt{2k^2-1}}\left(K(k)-\frac{1-k^2}{k^2} \Pi\left(\frac{2k^2-1}{k^2},k\right)\right)}\right)$ evaluated at $k=k_c$.}.
For $\Delta \phi$ exceeding this value, the disconnected graph is preferred, and the transition is found to be first order, as in the flat Minkowski space case \cite{Zarembo:1999bu}.  The flat Minkowski space limit is easy to read off, taking a $\theta_0\rightarrow 0$ limit, this reads $\Delta x_c = 0.9052 R$ agreeing with the result given in \cite{Zarembo:1999bu}.  We also find that for sufficiently large Wilson loops that the connected solution always dominates, due to the finite separation distance possible on the sphere.  This critical size is given approximately by
\bea
\sin^2(\theta_{\rm crit})=1-\frac{({\mathcal{E}}-1)^2}{({\mathcal{E}}+1)^2} \nn \\
\theta_{\rm crit}\approx 0.3547 \pi.
\eea
We see no problem with extending this to non-synchronous and non-equal-size Wilson loops, given the structure of the equations we find in the subsequent sections, although the above mentioned situation is the most symmetric.  We comment on the structure of these more general surfaces in the text and in the appendix.

One final note is in order.  When the disconnected configuration is preferred, we expect that the correct $\frac{1}{N^2}$ correction is given by the disconnected area (i.e. $1$, given that we have divided by this) times a propagation of a bulk field between the two disconnected caps: this is outside the scope of our current work, and has already been addressed in \cite{Berenstein:1998ij} for Minkowski space (for similar work for the null polygonal Wilson loops, see \cite{Alday:2010ku}).

We organize the paper as follows.  In section \ref{compute} we compute the equations of motion for the minimal surface in AdS$_n$, and write the two dimensional (2D) Lax pair associated with this system.  We reduce this Lax pair as in \cite{Arutyunov:2003za} and find a reduced one dimensional (1D) Lax pair.  This furnishes a set of 1D first order equations of motion, which we separate, and show how to solve by iterative integration.  In section \ref{solutions} we write down several explicit solutions, including that for the synchronous coaxial Wilson loops, and synchronous coaxial equal-sized Wilson loops.  We compute the regulated area minus the regulated area of the disconnected worldsheet to find the function $f_1$ above, and compute the Gross-Ooguri transition point, and the critical radius where the transition no longer occurs.  In section \ref{discussion} we discuss possible future work.


\section{Equations of motion, reduction to 1D system.}

\label{compute}


\subsection{Equations of motion, and reduction}


We first obtain the equations of motion for a string in AdS$_n$.  Recall that AdS$_n$ is given by the restriction
\be
-\left(Y^{-1}\right)^2-\left(Y^{0}\right)^2+\left(Y^{1}\right)^2\cdots +\left(Y^{n-1}\right)^2=-1 \label{quadconstraint}
\ee
in ${\mathbb R}^{2,n-1}$.  We will avoid using the flat space ${\mathbb R}^{2,n-1}$ metric for computing the scalar product, and simply write ``$\cdot$''.  We implement the above restriction via a Lagrange multiplier $\Xi$ in the Polyakov action, giving
\be
S=\int dz d\zb \left(\pa Y \cdot \pab Y - \Xi\left(Y\cdot Y+1\right)\right).
\ee
Upon elimination of $\Xi$, the equations of motion read
\bea
\pa \pab Y^M &=& (\pa Y \cdot \pab Y) Y^M  \nn \\
Y\cdot Y &=& -1 \label{eomadsA}
\eea
and further, we have the Virasoro constraint
\be
\pa Y \cdot \pa Y = \pab Y \cdot \pab Y =0. \label{viraads}
\ee
One may take the above (\ref{eomadsA}) and antisymmetrize in a $Y^N$ to find
\be
Y^{[N}\pa \pab Y^{M]} = 0. \label{eomcurrent}
\ee
This is in fact invertible, one may obtain (\ref{eomadsA}) from (\ref{eomcurrent}) by dotting in a $Y^N$, and using $Y\cdot Y=-1$.  Finally, we rewrite the above as
\be
\pa \left(Y^{[N} \pab Y^{M]}\right)+\pab \left(Y^{[N} \pa Y^{M]}\right) = 0 \label{eomcurrent2}
\ee
to write the above as a total divergence.  We can repackage these equations into equations of motion resulting from a Lax pair formulation, which we turn to now.

First, writing the fields in some matrix $g$ (which we explicitly show below), and defining $A=\pa g g^{-1}, \Ab=\pab g g^{-1}$, the equations of motion can be obtained via the Lax pair with spectral parameter $\lambda$
\be
\left(\pa-\frac{A}{1+\lambda}\right)\Psi=0, \qquad \left(\pab-\frac{\Ab}{1-\lambda}\right)\Psi=0.\label{laxpairA}
\ee
The integrability condition for the two equations is
\be
-\left(\frac{(1-\lambda)\pab A-(1+\lambda)\pa \bar{A}+[A,\bar{A}]}{(1+\lambda)(1-\lambda)}\right)\Psi=0.
\ee
which in the current context implies both the equation of motion (see footnote \ref{footnoteEOM})
\be
\pab A+\pa \bar{A}=0 \label{matrixEOM}
\ee
and gives the identity
\be
\pab A - \pa \bar{A}+[A, \bar{A}]=0 \label{laxid}.
\ee
In the above $A, \bar{A}$ plays the role of a gauge connection, and so we see that (\ref{laxpairA}) is invariant under
\bea
\Psi&=&\chi \Psi' \nn \\
A &=& \chi A' \chi^{-1}+ (1+\lambda)\pa \chi \chi^{-1} \nn \\
\Ab&=& \chi \Ab' \chi^{-1}+ (1-\lambda)\pab \chi \chi^{-1}. \nn \\
\eea
Usually, we allow for any transformation $\chi(z,\zb,\lambda)$ such that $A'$ and $\Ab'$ are again independent of $\lambda$.  However here, our concerns will be more general and so below we will allow for any $\chi$.

Next, we wish to write these equations in Cartesian components, and for this, we recall that $\pa_{\sigma}=\frac12(\pa + \pab), \pa_{\tau} = \frac12 i(\pa-\pab)$.  Therefore, in these coordinates, we find the Lax pair
\be
\left(\pa_{\sigma}-\frac12\left(\frac{A}{1+\lambda}+\frac{\Ab}{1-\lambda}\right)\right)\Psi=0, \qquad \left(\pa_{\tau}-\frac12i\left(\frac{A}{1+\lambda}-\frac{\Ab}{1-\lambda}\right)\right)\Psi=0.
\ee
which can be written
\be
\left(\pa_{\sigma}-\left(\frac{A_{\sigma}+i\lambda A_{\tau}}{1-\lambda^2}\right)\right)\Psi=0, \qquad
\left(\pa_{\tau}-\left(\frac{A_{\tau}-i\lambda A_{\sigma}}{1-\lambda^2}\right)\right)\Psi=0,
\ee
where $A_{\sigma}=\pa_{\sigma} g g^{-1}, A_\tau= \pa_{\tau} g g^{-1}$.  The gauge transformations can be read from the last ones quite easily and give
\bea
\Psi&=&\chi \Psi' \nn \\
A_{\sigma} &=& \chi A_{\sigma}' \chi^{-1}+ \pa_{\sigma} \chi \chi^{-1} -i\lambda \pa_{\tau}\chi \chi^{-1} \nn \\
A_{\tau}&=&  \chi A_{\tau}' \chi^{-1}+ \pa_{\tau} \chi \chi^{-1} +i \lambda \pa_{\sigma} \chi \chi^{-1}. \nn \\
\eea

We define the combination $\left(\frac{A_{\sigma}+i\lambda A_{\tau}}{1-\lambda^2}\right)=V_{\sigma}$ and similarly for $V_{\tau}$.  We will wish to reduce this model on a circle for a particular Ansatz, and we turn to this now.  A particular choice of $g$ for AdS$^5$ is given by \footnote{\label{footnoteEOM} Here we could have used the choice $g=Y\cdot \Gamma$ where $\Gamma^{I}$ are the $SO(2,n-1)$ (n=5) gamma matrices.  This is particularly easy because $\pa g g^{-1}= \pa Y_{I} Y_J \Gamma^{IJ}$, and likewise for the $\pab$ part of the connection.  This makes the equation (\ref{matrixEOM}) read exactly as (\ref{eomcurrent2}), so this is indeed the equations of motion.  Here, to reduce the size of the matrices we are working with the $4\times 4$ chiral components of the $SO(2,4)$ matrices, which are a constant similarity transformation away from the $g$ given here.}
\be
g=\begin{pmatrix} 0 & Z_1 & -Z_3 & \Zb_2 \\ -Z_1 & 0 & Z_2 & \Zb_3 \\ Z_3 & -Z_2 & 0 & -\Zb_1 \\ -\Zb_2 & -\Zb_3 & \Zb_1& 0 \end{pmatrix}
\ee
where we have defined
\bea
Z_1= Y^{-1} + i Y^{0}=e^{it}\sqrt{1+r^2}, &\quad& Z_2= Y^{1}+i Y^{2}=e^{i\phi}\cos(\theta)r, \nn \\
Z_3& =& Y^{3}+i Y^{4}=e^{i \psi} \sin(\theta)r.
\eea
This gives a particular parametrization of the homogeneous coordinates $Y^I$ that solves the quadratic constraint (\ref{quadconstraint}).  For this choice, the metric appears as
\bea
ds^2&=&\eta_{IJ}dY^I dY^J \nn \\
&=& -(1+r^2)dt^2+\frac{dr^2}{(1+r^2)}+r^2\left(d\theta^2+\cos^2(\theta)d\phi^2+\sin^2(\theta)d\psi^2\right)
\eea

We wish to make a reduction along a circle, and take that the $\sigma$ direction is compact, and we want to consider the solutions given by
\be
\psi(\sigma,\tau)=w \sigma
\ee
and all other functions are functions of $y$ only.  Now, we ask the question of whether there exists some connection $V_{\sigma}$ that is $\sigma$ independent after a gauge transformation.  Indeed we find that there is one, and it is given by
\be
\chi=\begin{pmatrix} e^{-\frac{i}{2} w \sigma} & 0 & 0 & 0 \\
                     0 & e^{\frac{i}{2} w \sigma}  & 0 & 0 \\
                     0 & 0 & e^{-\frac{i}{2} w \sigma} & 0  \\
                     0 & 0 & 0 & e^{\frac{i}{2}w \sigma} \end{pmatrix}.
\ee
In fact, this makes both $A_{\sigma}$ and $A_{\tau}$ independent of $\sigma$ altogether.  One may be concerned that the functions above are not well defined on the circle (e.g. $w$ an odd integer it is not well defined).  However, going around the circle in $\sigma$ either takes $\chi$ to itself or $\chi$ to minus itself, but in the gauge transformation, $\chi$ only appears ``squared'' so that the resultant connection is well defined.  The new connections $A_{\sigma}$ and $A_{\tau}$ depend on $\lambda$ in some new way, however.  Now we note that this simplifies our lives quite a bit.  The Lax equations now read
\be
\pa_{\tau} V_{\sigma} - \pa_{\sigma} V_{\tau} + [V_{\sigma}, V_{\tau}]=0
\ee
however, $V_{\tau}$ is $\sigma$ independent, and so we find
\be
\pa_{\tau} V_{\sigma} = [V_{\tau}, V_{\sigma}]
\ee
which can be thought of as the one dimensional Lax pair
\be
L'=[M,L]
\ee
where $'$ is now $\tau$ differentiation.
In such a formulation, the conserved quantities are given by $\Tr(L^n)$ because $\pa_{\tau} \Tr(L^n)=n\Tr(L'L^{n-1})=n\Tr([M,L]L^{n-1})=0$ by the cyclicity of the trace.  Since $L$ is still a rational function of $\lambda$, we expect there to be only a finite number of conserved quantities of this form.  The above discussion is in fact the same reduction as was found in \cite{Arutyunov:2003za}: here we are using a subclass of Ansatz setting certain winding numbers to zero, and setting some dynamical functions to zero.  However, here we will be considering the most general solutions to the system that we have found, rather than focusing on ``constant radii'' solutions discussed in \cite{Arutyunov:2003za}.

In fact, the above Lax pair is insufficient to furnish all conserved quantities.  However, it does furnish one that is not so obvious.  To obtain a Lax pair furnishing all conserved quantities, one may simply redefine $L+I(P_t+\lambda P_\phi)$ where $I$ is the identity matrix, and $P_t$ and $P_\phi$ are conserved quantities that we can find some other way\footnote{Perhaps there is a more ``efficient'' Lax pair than the one we find here}.  This will not be needed for our further discussion, and we simply state a set of conserved quantities
\bea
&& P_t= (1+r^2)t' \nn \\
&& P_\phi= r^2\cos^2(\theta)\phi' \nn \\
&& H_1= -(1+r^2)(t')^2+\frac{(r')^2}{1+r^2}+r^2\left((\theta')^2+\cos^2(\theta)(\phi')^2-w^2\sin^2(\theta)\right) \\
&& H_2= \sin^2(\theta)(1+r^2)r^2 (t')^2+\frac{(r')^2 \sin^2(\theta)}{1+r^2}+2\sin(\theta)\cos(\theta)r \theta' r' \nn \\ && +\left(\cos^2(\theta)-r^2\sin^2(\theta)\right)r^2(\theta')^2 +r^2\left(-\sin^2(\theta)\cos^2(\theta)r^2(\phi')^2-w^2\sin^2(\theta)\right).
\nn
\eea
Note that $H_1$ is just the the hamiltonian resulting from the reduced action.  Further, it must be $0$ for the above system to correspond to a minimal surface: $H_1=0$ is the Virasoro constraint.

Next, we will eliminate $t$ and $\phi$ from the discussion by replacing them with the conserved quantities $P_t$ and $P_\phi$.  Doing so, we find
\bea
&& H_1= -\frac{P_t^2}{1+r^2}+\frac{(r')^2}{1+r^2}+r^2\left((\theta')^2+\frac{P_\phi^2}{r^4\cos^2(\theta)}-w^2\sin^2(\theta)\right) \\
&& H_2= \frac{P_t^2 r^2 \sin^2(\theta)}{1+r^2} +\frac{(r')^2 \sin^2(\theta)}{1+r^2}+2\sin(\theta)\cos(\theta)r \theta' r' \nn \\ && +\left(\cos^2(\theta)-r^2\sin^2(\theta)\right)r^2(\theta')^2 +r^2\left(-\frac{P_\phi^2 \sin^2(\theta)}{r^2\cos^2(\theta)}-w^2\sin^2(\theta)\right).
\eea
The above quantities are conserved simply using the reduced equations of motion.  However, since we are looking for a particular class of solutions, $H_1=0$, we will find other linear combinations more useful.  We take instead the combinations
\bea
\mcH_1&=&(H_2-H_1)=-\frac{\left(-\cos(\theta)r'+\sin(\theta)r(1+r^2)\theta'\right)^2}{1+r^2} \nn \\
&& \qquad \qquad \qquad -\frac{(1+r^2\sin^2(\theta))((1+r^2)P_\phi^2-P_t^2r^2\cos^2(\theta))}{r^2(1+r^2)\cos^2(\theta)} \\
\mcH_2 &=& (H_2+r^2\sin^2(\theta)H_1)=\left(\sin(\theta)r'+r\cos(\theta)\theta'\right)^2\nn \\
&& \qquad \qquad \qquad
-w^2r^2\sin^2(\theta)\left(1+r^2\sin^2(\theta)\right)
\eea
where now the Virasoro constraint reads $\mcH_1=\mcH_2\equiv \mcH$.  Note that $\mcH_2$ is only conserved quantities when the Virasoro constraint is met.  Further note that the above combinations are special in that the single derivative term squared can be written as a total derivative.  This allows us to write the above differential equations as
\bea
&&\pa_\tau \left(\frac{r\cos(\theta)}{\sqrt{1+r^2}}\right)= \nn \\ &&\qquad  \pm \frac{1}{1+r^2}\sqrt{-\mcH+P_\phi^2+P_t^2-P_t^2\left(\frac{r\cos(\theta)}{\sqrt{1+r^2}}\right)^2 -P_\phi^2\frac{1}{\left(\frac{r\cos(\theta)}{\sqrt{1+r^2}}\right)^2}} \\
&& \pa_\tau \left(r\sin(\theta)\right)=\pm \sqrt{w^2(r\sin(\theta))^4+w^2(r\sin(\theta))^2+\mcH}.
\eea
The second equation can be solve directly by integrating (note the right hand side is a function of $r\sin(\theta)$ only).  Let us denote
\be
F_1=\frac{r\cos(\theta)}{\sqrt{1+r^2}}, \qquad F_2=r\sin(\theta),
\ee
so that the above equations may be written
\bea
F_2'=\pm \sqrt{w^2 F_2^4+w^2 F_2^2+ \mcH} \label{F2EOM} \\
F_1'=\pm \frac{1-F_1^2}{1+F_2^2}\sqrt{-\mcH+P_\phi^2+P_t^2-P_t^2F_1^2 -P_\phi^2\frac{1}{F_1^2}}. \label{F1EOM}
\eea
Along with the equations
\bea
&& t'= P_t\frac{1-F_1^2}{1+F_2^2} \label{tEOM} \\
&& \phi'=P_\phi \frac{1-F_1^2}{F_1^2(1+F_2^2)}\label{phiEOM}
\eea
give all of the equations of motion.  Therefore, one solves (\ref{F2EOM}) for $F_2$, then using this result, solves (\ref{F1EOM}).  This is possible because the right hand side of (\ref{F1EOM}) is a function of $F_1$ times a function of $F_2$: since $F_2$ is a known function of $\tau$, one can integrate this with respect to $\tau$ and integrate with respect to $F_1$ on the left, furnishing $F_1$ as a function of $\tau$.  With $F_1$ and $F_2$ in hand, one may then directly integrates (\ref{tEOM}) and (\ref{phiEOM}) find the general result.  There is an important simplification here.  When integrating the $F_1$ equation, (\ref{F1EOM}), one will be forced to do the integral $\int \frac{d\tau}{1+F_2^2}$, and then $F_1$ will be written as a function of this variable.  However, this measure $\frac{d\tau}{1+F_2^2}$ will appear when integrating to find $t$ and $\phi$ as well.  Therefore, one can change to the coordinates defined by this integration measure, and simplify the integration.  We will use this trick later, and emphasize it again when we turn to computing the solutions.

Finally, if one wishes, one can solve the equations for non zero $H_1$, and this only changes the coefficient of $F_2^2$ in (\ref{F2EOM}), and the value of $\mcH$ appearing in (\ref{F1EOM}), and so does not affect the method of solving the equations, nor the form of the equations.  When considering profiles on the S$^5$ part of the geometry this will become important, but does not appear to be a major obstacle, given these simple modifications (see \cite{Tseytlin:2002tr} for example).

We now segue a bit and discuss the sections of $F_1$ and $F_2$ that are allowed.  First, $r$ is a positive definite number, and so the signs of $F_1$ and $F_2$ are actually given by $\theta$.  Recall that for one cover of the S$^3$ inside of AdS$_5$ we have that $0\leq \theta \leq \frac{\pi}{2}$ to get a single covering.  Extending $\theta$ beyond this bound is relatively straightforward, as passing through $\theta=0$ can be compensated for by $(-\theta,\phi,\psi)\rightarrow (\theta,\phi,-\psi)$, as these define the same homogeneous coordinates (and is the usual way for going through such an ``origin'').  Similarly, for going through $\theta=\frac{\pi}{2}$ we see that $\left(\frac{\pi}{2}+\theta,\phi,\psi\right)\rightarrow \left(\frac{\pi}{2}-\theta,-\phi,\psi\right)$.  We can, therefore, make sense out of all signs for $F_1$ and $F_2$.  However, note that $-1<F_1=\frac{r\cos(\theta)}{\sqrt{1+r^2}}<1$.  Hence, any solution that we find for $F_1$ must lie in this range.  Another way of saying this is that
\be
r^2=\left(\frac{1+F_2^2}{1-F_1^2}-1\right),\qquad \tan^2(\theta)=\frac{F_2^2}{F_1^2}\frac{1-F_1^2}{1+F_2^2}.
\ee
One should note that these expression define a {\it real} $r$ and a {\it real} $\theta$ for any real $F_1,F_2$ that satisfy $F_1^2\leq 1$, and no restrictions on $F_2$.

There is one final note.  Given the Virasoro constraint, one can get the pullback metric immediately as
\be
ds^2=w^2F_2^2\left(d\tau^2+d\sigma^2\right) \label{pullbackgen}
\ee
allowing direct computation of the lagrangian density as
\be
S=\frac{L^2w^2}{4\pi\alpha'}\int d\tau d\sigma F_2^2
\ee
where we have reintroduced the AdS radius $L$.  The above should be read with some care, however, when talking about Euclidean worldsheets: these require an extra factor of $i$.  This will render our exponentiated action as being exponential suppression, and so should be understood as a tunneling event: we have already accounted for this in equation (\ref{introVev}), where the area is to be read as the classical Euclidean area of the worldsheet.  This exponential suppression is not surprising because we will be considering space-like separated \footnote{All points of each Wilson loop are space-like separated from all points of the other Wilson loop} space-like Wilson loops.


\section{Solutions}
\label{solutions}



\subsection{Special case: $\mcH=0$}


Let us work out a simpler set of equations as a warm up.  We take $\mcH=0$ above, and take $\pm=+$ for the $F_1$ equation and $\pm=-$ for the $F_2$ equation, and find
\bea
&& F_2(\tau)=\frac{1}{\sinh(w(\tau-\tau_0))} \\
&& F_1(\tau)=\sqrt{\frac{(P_t^2-P_\phi^2)^2\left(C_2+(\tau-\tau_0)-\frac{1}{w}\tanh(w(\tau-\tau_0))\right)^2+P_\phi^2}{ (P_t^2-P_\phi^2)^2\left(C_2+(\tau-\tau_0)-\frac{1}{w}\tanh(w(\tau-\tau_0))\right)^2+P_t^2}}.
\eea
Note that this $F_1$ and $F_2$ define a good $\theta$ and $r$ given the restrictions mentioned above, and further, the signs were chosen so that $0\leq \theta \leq \frac{\pi}{2}$.  Particularly, $F_1^2<1$ is always true as long as $P_t^2>P_\phi^2$, i.e. that $|P_t|>|P_\phi|$.

Given the above functions, one can integrate the expressions for $t(\tau)$ and $\phi(\tau)$ and further, one may rearrange the above expressions to solve for $r(\tau)$ and $\theta(\tau)$.  Upon doing so, we find
\bea
&& r(\tau)=\sqrt{\frac{\cosh^2(w (\tau-\tau_0))\left((P_t^2-P_\phi^2) g_0(\tau)^2+\frac{P_\phi^2}{P_t^2-P_\phi^2}+\frac{1}{\cosh^2(w (\tau-\tau_0))}\right)}{\sinh^2(w (\tau-\tau_0))}} \label{K0r}\\
&& \theta(\tau)=\arctan\left(\sqrt{\frac{P_t^2-P_\phi^2}{\cosh^2(w (\tau-\tau_0))\left((P_t^2-P_\phi^2)^2g_0(\tau)^2+P_\phi^2\right)}}\right) \label{K0th} \\
&& t(\tau)=\arctan\left(\frac{(P_t^2-P_\phi^2)g_0(\tau)}{P_t}\right) + t_0 \label{K0t}\\
&& \phi(\tau)=\arctan\left(\frac{(P_t^2-P_\phi^2)g_0(\tau)}{P_\phi}\right) + \phi_0 \label{K0ph}
\eea
where we have defined the useful functions
\be
g_0(\tau)=C_2+(\tau-\tau_0)-\frac{1}{w}\tanh(w(\tau-\tau_0)).
\ee
Note that in the above, we have the following constants of motions $P_t, P_\phi, C_2, t_0, \phi_0, \tau_0$, and we have 4 dynamical fields.  In general, we would expect 8 constants of motion for these equations, however, we must take $H_1=0$ to describe minimal surfaces, and we have taken the slice $\mcH=0$ as well, hence we are left with six constants describing the surface.  Finally, note that in the limit $\tau\rightarrow 0$ that the surface goes to a circle, and $\tau\rightarrow \infty$ it goes to a point.  This may have some bearing on the calculations performed in \cite{Arutyunov:2001hs}, however, we do not regulate these areas, and so we will have little to say on this point here.


\subsubsection{Sub-sub case: the single cap solution.}


We now explore a particular solution with $F_1=$ const, $\mcH=0, w=1, P_t=P_\phi=0$ that will be important for later.  Note that in the above discussion, plugging in $\mcH=0$, one can take a limit with $F_1^2 \rightarrow P_\phi^2/P_t^2$ being a fixed ratio less than one, but still get $P_\phi=P_t=0$.  In this limit, the solution reduces to
\bea
F_2=\frac{1}{\sinh{\tau}} \nn \\
F_1= F_0\equiv \cos(\theta_0) \nn \\
t=t_0 \nn \\
\phi=\phi_0.
\eea
where we have taken $w=1$ for a single winding.  For multiple windings, one can simply take $\tau\rightarrow w \tau$ in the above equation.  This ``single cap'' solution is well known \cite{Drukker:2000rr,Berenstein:1998ij}

This gives a particularly simple solution
\bea
r^2=\frac{\cos^2(\theta_0)+\frac{1}{\sinh^2(\tau)}}{\sin^2(\theta_0)} \\
\theta=\arctan\left(\frac{\tan(\theta_0)}{\cosh(\tau)}\right).
\eea
This solution is the ``single cap'' solution.  This is easiest to see in the following way.  First, note that when $\tau\rightarrow 0$, the radius goes to infinity, and theta goes to some value fixed by $\theta_0$.  Therefore, the constant $\sin(\theta_0)$ determines the radius of the Wilson loop.  Then, when $\tau\rightarrow \infty$ we see that $\theta\rightarrow 0$ and $r\rightarrow$ const, where again this constant is determined by $\theta_0$.  However, we note that the pullback metric (\ref{pullbackgen}) gives
\be
ds^2=\frac{1}{\sinh^2(\tau)} \left(d\tau^2+d\sigma^2\right).
\ee
We would like to determine what the nature of the ``cutoff'' is when $\tau\rightarrow \infty$.  It is easy to see that in this limit, the pullback metric (\ref{pullbackgen}) reduces to
\be
ds^2=e^{-2\tau} \left(d\tau^2+d\sigma^2\right),
\ee
and changing variables to $T=e^{-\tau}$, we see that the metric becomes
\be
ds^2=\left(dT^2+T^2 d\sigma^2\right),
\ee
where now the above limit is $T\rightarrow 0$.  Because the periodicity of $\sigma$ is $2\pi$, this represents a smooth point of the worldsheet, and is just a place where the round ``cap'' surface pinches off to zero size in a smooth way.  If we had instead left $w$ to be an integer winding number, the asymptotic form of the metric would have become
\be
ds^2=w^2e^{-2w \tau}\left(d\tau^2+d\sigma^2\right)
\ee
which can be transformed by the coordinate transformation $T=e^{-w \tau}$
\be
ds^2=dT+w^2T^2d\sigma^2.
\ee
This gives a conical singularity at the point $T=0$, with an excess of angular ``space'' given by $w$.  This, of course, is easy to visualize: the multiple windings are collapsing to a point, giving multiple covers of the flat space metric.


\subsection{Finding $F_2$ for general conserved quantities.}


Now let us look to the general $\mcH\neq 0$ case.  We write this as
\be
(F_2')^2=(w^2 F_2^4+w^2 F_2^2+ \mcH)=\mcH\left(1-\frac{F_2^2}{-\frac12+\frac{\sqrt{1-\ga}}{2}}\right) \left(1-\frac{F_2^2}{-\frac12-\frac{\sqrt{1-\ga}}{2}}\right)
\ee
where
\be
\ga=\frac{4\mcH}{w^2}.
\ee
We recognize the above differential equation as being solve by Jacobi elliptic functions, however, it is not immediately clear how to write them in a manifestly real way.  However, it is clear is to break the discussion into cases where certain reality conditions are met.  We note that the combination $\sqrt{1-\ga}$ is either real or imaginary depending on whether $\ga$ is larger or smaller than 1 ($w$ is assumed to be real).  Hence, $\ga=1$ is a special value (in fact, this makes the right hand side of the equation above a perfect square, and this can be solved nicely too).  However, there is also a crossover when $-\frac12+\frac{\sqrt{1-\ga}}{2}$ is positive or negative, in other words when $\ga=0$ (this is the case $\mcH=0$ above).  Hence we wish to break the problem into three cases:
\begin{enumerate}
\item $\ga<0$ ($\mcH<0$)
\item $0<\ga<1$ ($0<\mcH<w^2/4$)
\item $\ga>1$ ($\mcH>w^2/4$).
\end{enumerate}

\begin{figure}
\centering
\includegraphics[width=0.4\textwidth]{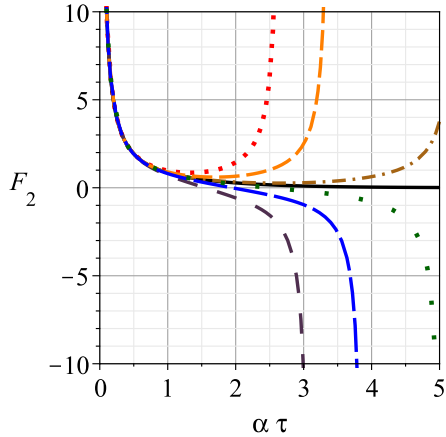}
\caption{Here we have plotted $F_2(\tau)$ for one connected component of the worldsheet (where the arguments of the elliptic functions range between $0$ and $2K(k)$ for the appropriate $k$).  The graphs depict different values of $\gamma=[-5,-2,-0.3,0,0.5,2,6]$ corresponding to the colors (line styles) [red (dot), orange (dash), brown (dash-dot), black (solid), dark green (wide space dot), blue (long dash), violet (wide spaced dash)].}
\label{F2graphs}
\end{figure}

After a bit of work, and using various identities \cite{Whit,Grad,NIST}, we can arrive at the following solutions for these cases
\be
\pm F_2=
\begin{cases}
(1-\ga)^{\frac14}\left[\frac{\dn}{\sn}\left(w (1-\ga)^{\frac14} \tau, \frac{1}{\sqrt{2}}\frac{\sqrt{\sqrt{1-\ga}+1}}{(1-\ga)^{\frac14}}\right)\right], & \mbox{if $\ga<0$, ($\mcH<0$)} \\
\sqrt{\frac12+\frac{\sqrt{1-\ga}}{2}}\left[\frac{\cn}{\sn}\left(\frac{\sqrt{\ga}}{2}\frac{w \tau}{\sqrt{\frac12-\frac{\sqrt{1-\ga}}{2}}},\frac{\sqrt{2}(1-\ga)^{\frac14}}{\sqrt{1+\sqrt{1-\ga}}}\right)\right]& \mbox{if $0<\ga<1$, ($0<\mcH<\frac{w^2}{4}$)} \\
\frac{\ga^{\frac14}}{\sqrt{2}}\left[\frac{\cn}{\sn \cdot \dn}\left(\frac{w \tau}{\sqrt{2}\sqrt{\ga}},\frac{\sqrt{\sqrt{\ga}-1}}{\sqrt{2}\ga^{\frac14}}\right)\right] & \mbox{if $\ga>1$, ($\mcH>\frac{w^2}{4}$)}
\end{cases}
\ee
where the relative ``phases'' associated with shifting $\tau\rightarrow \tau-\tau_0$ have been chosen so that the terms agree on the boundary of their relative regimes ($\ga=0,1$).  Above we have used a condensed notation where it is understood that all elliptic function have the same arguments, for example
\be
\left[\frac{\cn}{\sn \cdot \dn}(z,k)\right]\equiv \frac{\cn(z,k)}{\sn(z,k) \dn(z,k)}.
\ee
Also note that in the above, the modulus (usually denoted ``$k$'') of all functions is less than one.  This gives some nice results.  First, because $0<k<1$ we have that $\dn$ is positive definite, and both $\cn$ and $\sn$ are functions with amplitude $1$.  Thus, $\frac{\dn}{\sn}$ is a generalization of a ``$\frac{1}{\sin}$'' function.  We can see this by remembering that all elliptic function have period $4K(k)$ where $K(k)$ is the complete elliptic function, and that $\sn(0,k)=0$ and $\sn(2K(k),k)=0$: it goes to zero at 0 and at the period over 2.  This then will have exactly the same shape as $\frac{1}{\sin}$.  Similarly, $\frac{\cn}{\sn}$ and $\frac{\cn}{\sn \cdot \dn}$ are in some sense generalizations of $\cot$ (blowing up at 0 and period over 2, but crossing the horizontal $\tau$ axis).  This gives a rough idea of how these functions behave.  We plot them to emphasize the point in figure \ref{F2graphs}.

We note that in these graphs where $\mcH>0$ that the worldsheet has been reflected through the origin: it has shrunk to zero size and then reemerged to finite size.  We take that this is an orientation reversal of the world sheet. and this corresponds to reversing the orientation of the second Wilson loop operator.  Hence, later when we define ``coaxial'' we will mean ``coaxial with the same orientation.''  Reversing the orientation of one Wilson loop is considering the complex conjugate of the original Wilson loop operator, and we expect the correlator to depend on this information.  We will not consider correlators of this form, ``$\langle W[C] W[C]^\dagger \rangle$'', and leave this to future work.


\subsection{Synchronous coaxial Wilson loops, and the Gross Ooguri phase transition.}


We now turn to looking at a special class of solutions given a pair of Wilson loop operators that are synchronous, spacelike separated, and with the same orientation.  Given these conditions and the above discussion, we see that this is a case where $P_t=0$, but $P_\phi\neq 0$ and $\mcH<0$ (see also appendix \ref{OtherConstantsF1}).  In this case, the equations of motion simplify, and rather than arctan functions for the $F_1$ integral, as in appendix \ref{OtherConstantsF1}, we get instead logs.  Solving these equations becomes relatively straightforward, as one simply needs to integrate to find $F_1$, and we find
\bea
&& F_2(\tau)=(1-\ga)^{\frac14}\left[\frac{\dn}{\sn}\left(w (1-\ga)^{\frac14} \tau, \frac{1}{\sqrt{2}}\frac{\sqrt{\sqrt{1-\ga}+1}}{(1-\ga)^{\frac14}}\right)\right], \quad \ga=\frac{4\mcH}{w^2}<0 \nn \\
&& F_1(\tau)^2=\frac{\left(e^{2\sqrt{-\mcH}\int \frac{d\tau}{1+F_2(\tau)^2}}+4P_\phi^4+4P_\phi^2\mcH\right)^2-64P_\phi^6\mcH}{\left(e^{2\sqrt{-\mcH}\int \frac{d\tau}{1+F_2(\tau)^2}}+4P_\phi^4-4P_\phi^2\mcH\right)^2} \\
&& t(\tau)=t_0 \\
&& \phi(\tau)=\int d\tau P_\phi\frac{1-F_1^2}{F_1^2(1+F_2^2)}.
\eea
We will now turn our attention to the $\frac{d\tau}{1+F_2^2}$ integral.  For this case, we find
\bea
&& \int\frac{d\tau}{1+F_2^2(\tau,k)} \nn \\
&=&\frac{1}{w(1-\gamma)^{\frac14}}\int \frac{dz}{1+F_2(z,k)^2} \\
&&=\frac{1}{w(1-\gamma)^{\frac14}}\Bigg[z-K(k)-\frac{(1-k^2)}{k^2}\Bigg( \Pi\left(\sn(z-K(k),k),\frac{2k^2-1}{k^2},k\right)\nn \\
&& \qquad \qquad \qquad \qquad \qquad \qquad \qquad \qquad \qquad \qquad +\Pi\left(1,\frac{2k^2-1}{k^2},k\right)\Bigg)\Bigg]\nn
\eea
where
\be
k=\frac{\sqrt{\frac{\sqrt{1-\gamma}}{2}+\frac12}}{(1-\gamma)^{\frac14}},\qquad z=w (1-\gamma)^{\frac14} \tau
\ee
and the definition of the elliptic pi function \footnote{Here we have chosen to take a certain type of elliptic $\Pi$ function so that we avoid branch cuts at $z=(2n+1)K(k)$, and instead have branch cuts at $z=2nK(k)$ to worry about.  We do this because the later set of branch cuts occur at points where the worldsheet goes to the boundary of AdS, rather than in some interior point.} is given in (\ref{defPi}).
The above answer is good enough for values of $z$ in the interval $(0,2K(k))$, which is all we will really need (this interval defines a single connected component of the solution that begins and ends on the boundary of AdS).
However, perhaps being a bit pedantic, we find a solution that is good for all values of $z$ on the real axis (for the values of $k$ needed).  For this, we must be more careful with the signs of various functions, as $\Pi(T,\alpha^2,k)$ has branch-cuts when $T=1$ for the values of $\alpha^2$ and $k$ given (both are less than one).  For example, one should note that $\Pi(\sn(z-K(k),k),\alpha^2,k)$ behaves as a constant plus a square root of $[1-\sn(z-K(k),k)]$ as $\sn(z-K(k),k)\rightarrow 1$: note that $\sn(z,k)$ goes to 1 quadratically, and so the function is linear in $z$ about this point.  However, the sign of the linear correction is ambiguous: this is because we are evaluating $\sqrt{x^2}$ which can be either $\pm x$.  Hence, we must put in a $\Theta(\sn(z,k))$ to keep track of signs, where $\Theta$ is a step function that is $+1$ for positive values and $-1$ for negative values of its argument:
\be
\Theta(x)=\begin{cases}1 & \mbox{if $x>0$} \\
{\mbox{undefined}} & \mbox{if x=0} \\
-1 & \mbox{if $x<0$}. \end{cases}
\ee
However, now we see that we have introduced step function discontinuities at $z=2mK(k)$ with integer $m$.  These are easily removed with a ``descending stairs'' function $-i\ln(e^{i x})-x$ for some $x$ given a branch cut for $\ln(z)$ that lies along the negative real axis $(-\infty,0)$.  Putting this together, we find the final answer
\bea
&&\frac{1}{w(1-\gamma)^{\frac14}}\int_0^{z} \frac{dz}{1+\left(\frac{\dn}{\sn}(z,k)\right)^2\frac{1}{2k^2-1}} \\
&&\qquad =\frac{1}{w(1-\gamma)^{\frac14}}\Bigg[z-\frac{(1-k^2)}{k^2} \Pi\left(\sn(z-K(k),k),\frac{2k^2-1}{k^2},k\right)\Theta(\sn(z,k)) \nn \\
&&\qquad \quad + \frac{(1-k^2)\Pi\left(1,\frac{2k^2-1}{k^2},k\right)}{k^2}\left(-\frac{i}{\pi}\ln\left(e^{i\pi\left(\frac{z}{K(k)}-1\right)}\right) -\frac{z}{K(k)}\right)\Bigg].\nn \\
&& I_1(\tau)\equiv \Bigg[z-\frac{(1-k^2)}{k^2} \Pi\left(\sn(z-K(k),k),\frac{2k^2-1}{k^2},k\right)\Theta(\sn(z,k)) \nn \\
&&\qquad \quad + \frac{(1-k^2)\Pi\left(1,\frac{2k^2-1}{k^2},k\right)}{k^2}\left(-\frac{i}{\pi}\ln\left(e^{i\pi\left(\frac{z}{K(k)}-1\right)}\right) -\frac{z}{K(k)}\right)\Bigg].
\eea
One should keep in mind that the earlier function is sufficient for our needs, as this will describe one ``boundary to boundary'' section of the string.

We have thus found the solution for $F_1$ given by
\bea
&& F_1(\tau)^2=\frac{\left(C_2 \exp\left({\frac{\sqrt{-\ga}}{(1-\ga)^{\frac14}}I_1}\right)+4P_\phi^4+4P_\phi^2\mcH\right)^2-64P_\phi^6\mcH}{\left(C_2 \exp\left({\frac{\sqrt{-\ga}}{(1-\ga)^{\frac14}}I_1}\right)+4P_\phi^4-4P_\phi^2\mcH\right)^2} \nn\\
&& I_1(\tau)\equiv \Bigg[z-\frac{(1-k^2)}{k^2} \Pi\left(\sn(z-K(k),k),\frac{2k^2-1}{k^2},k\right)\Theta(\sn(z,k)) \nn \\
&&\qquad \quad + \frac{(1-k^2)\Pi\left(1,\frac{2k^2-1}{k^2},k\right)}{k^2}\left(-\frac{i}{\pi}\ln\left(e^{i\pi\left(\frac{z}{K(k)}-1\right)}\right) -\frac{z}{K(k)}\right)\Bigg].\nn \\
&& k=\frac{\sqrt{\frac{\sqrt{1-\gamma}}{2}+\frac12}}{(1-\gamma)^{\frac14}},\qquad z=w (1-\gamma)^{\frac14} \tau, \qquad C_2>0
\eea
and one takes the positive square root to determine $F_1$.  We have included an ``integration constant'' $C_2$ here, associated with the fact that we have defined $I_1$ by a definite integral.  This $F_1$ lies in the acceptable range $0<F_1^2<1$.  This can be seen by noting that the numerator and denominator are both positive definite (given that $\mcH<0$), and noting that the numerator minus the denominator is sign definite, given the sign of $C_2$, and requires $C_2>0$ so that it is negative.  This restriction on $C_2$ is quite natural because it comes about as the exponentiation of an integration constant: if this constant is real, $C_2>0$.  Finally, because $F_2$ is real, these functions define a real $r$ and $\theta$.

The final integration for $\phi$ can be explicitly performed.  This is because $F_1$ is a function of the variable $\int \frac{\d\tau}{1+F_2^2}$.  The $\phi$ integral actually has this factor $\frac{d\tau}{1+F_2^2}$ built in, and so we integrate this directly, obtaining\footnote{note that if we had let $P_t\neq0$ that the $t$ equation of motion shares this property, so in the general case, both of these equations may be integrated using this variable to eliminate $F_2$ in a straightforward way.}
\be
\phi=\phi_0 - \arctan\left(\frac{C_2\exp\left({\frac{\sqrt{-\ga}}{(1-\ga)^{\frac14}}I_1}\right) +4P_\phi^4+4P_\phi^2\mcH}{8P_\phi^3\sqrt{-\mcH}}\right).
\ee
So far, our discussion has been for general radius Wilson loops.  In fact, one can see immediately that since $F_1$ is in the appropriate range, and because $I_1$ has no singularities at finite $\tau$, that $F_1$ goes to constant values when $F_2$ diverges.  This gives that the endpoints of $0<\tau<\frac{2K(k)}{w(1-\ga)^\frac14}$ are at the boundary of AdS (this is the only way to get a divergent $F_2$ and finite $F_1$).  In fact, at these endpoints, $C_2$ helps fix the size of the two Wilson loops.

Next, we note that the function $I_1$ is monotonically increasing for increasing $\tau$: this is easy to understand because it is related to the integral $\int\frac{d\tau}{1+F_2^2}$ which has a positive definite integrand.  Therefore we see that there is no way for the function $g=C_2 \exp\left({\frac{\sqrt{-\ga}}{(1-\ga)^{\frac14}}I_1}\right)$ to be constant unless $C_2=0$, but this case is $F_1=1$ which is forced to be on the boundary the entire time with $\theta=0$ fixed.  However, we may ask the question of when $F_1$ has the same value at the two endpoints, namely when $F_1(\tau=0)=F_1\left(\tau=\frac{K(k)}{w(1-\ga)^\frac14}\right)$: this is where the two Wilson loops have the same size.  This occurs when
\bea
&&\frac{
\left(C_2 \exp\left({\frac{\sqrt{-\ga}}{(1-\ga)^{\frac14}}I_1(0)}\right) +4P_\phi^4+4P_\phi^2\mcH\right)^2-64P_\phi^6\mcH}{ \left(C_2\exp\left({\frac{\sqrt{-\ga}}{(1-\ga)^{\frac14}}I_1(0)}\right) +4P_\phi^4-4P_\phi^2\mcH\right)^2} \nn \\
&& -
\frac{\left(C_2 \exp\left({\frac{\sqrt{-\ga}}{(1-\ga)^{\frac14}}I_1\left(\frac{K(k)}{w(1-\ga)^\frac14}\right)}\right) +4P_\phi^4+4P_\phi^2\mcH\right)^2-64P_\phi^6\mcH}{ \left(C_2\exp\left({\frac{\sqrt{-\ga}}{(1-\ga)^{\frac14}}I_1\left(\frac{K(k)}{w(1-\ga)^\frac14}\right)}\right) +4P_\phi^4-4P_\phi^2\mcH\right)^2}=0.
\eea
This happens when either
\be
C_2 \exp\left({\frac{\sqrt{-\ga}}{(1-\ga)^{\frac14}}I_1(0)}\right)= C_2\exp\left({\frac{\sqrt{-\ga}}{(1-\ga)^{\frac14}}I_1\left(\frac{K(k)}{w(1-\ga)^\frac14}\right)}\right)
\ee
which, again, due to $I_1$ being monotonic, cannot happen unless $C_2=0$; or they are equal when
\be
C_2 \exp\left({\frac{\sqrt{-\ga}}{(1-\ga)^{\frac14}}I_1(0)}\right)=\frac{16P_\phi^4\left(P_\phi^2-\mcH\right)^2} {C_2\exp\left({\frac{\sqrt{-\ga}}{(1-\ga)^{\frac14}}I_1\left(\frac{K(k)}{w(1-\ga)^\frac14}\right)}\right)}.
\ee
We have normalized so that $I_1(0)=0$ and so we get
\be
C_2=\frac{4P_\phi^2\left(P_\phi^2-\mcH\right)}{\exp\left({\frac{\sqrt{-\ga}} {2(1-\ga)^{\frac14}}I_1\left(\frac{K(k)}{w(1-\ga)^\frac14}\right)}\right)}.
\ee
This value of $C_2$ is special.  It essentially dresses the factor of $\exp\left({\frac{\sqrt{-\ga}}{(1-\ga)^{\frac14}}I_1(\tau)}\right)$ with a factor $4P_\phi^2\left(P_\phi^2-\mcH\right)$ and re-zeros the function $I_1$ such that it is odd via reflection about $\tau=\frac{K(k)}{w(1-\ga)^\frac14}$.  This, in fact, renders the function $F_1$ even with respect to this reflection.

Therefore, the solution connecting two synchronous coaxial equal radius Wilson loops is
\bea
&& F_2(\tau)=(1-\ga)^{\frac14}\left[\frac{\dn}{\sn}\left(z, k\right)\right] \nn \\
&& F_1(\tau)^2=\frac{\left(C_2 \exp\left({\frac{\sqrt{-\ga}}{(1-\ga)^{\frac14}}I_1}\right)+4P_\phi^4+4P_\phi^2\mcH\right)^2-64P_\phi^6\mcH}{\left(C_2 \exp\left({\frac{\sqrt{-\ga}}{(1-\ga)^{\frac14}}I_1}\right)+4P_\phi^4-4P_\phi^2\mcH\right)^2} \nn\\
&& \phi=\phi_0 - \arctan\left(\frac{C_2\exp\left({\frac{\sqrt{-\ga}}{(1-\ga)^{\frac14}}I_1}\right) +4P_\phi^4+4P_\phi^2\mcH}{8P_\phi^3\sqrt{-\mcH}}\right). \nn \\
&& C_2=\frac{4P_\phi^2\left(P_\phi^2-\mcH\right)}{\exp\left({\frac{\sqrt{-\ga}} {2(1-\ga)^{\frac14}}I_1\left(\frac{K(k)}{w(1-\ga)^\frac14}\right)}\right)}\nn \\
&& I_1(\tau)\equiv \Bigg[z-\frac{(1-k^2)}{k^2} \Pi\left(\sn(z-K(k),k),\frac{2k^2-1}{k^2},k\right)\Theta(\sn(z,k)) \nn \\
&&\qquad \quad + \frac{(1-k^2)\Pi\left(1,\frac{2k^2-1}{k^2},k\right)}{k^2}\left(-\frac{i}{\pi}\ln\left(e^{i\pi\left(\frac{z}{K(k)}-1\right)}\right) -\frac{z}{K(k)}\right)\Bigg].\nn \\
&& k=\frac{\sqrt{\frac{\sqrt{1-\gamma}}{2}+\frac12}}{(1-\gamma)^{\frac14}},\qquad z=w (1-\gamma)^{\frac14} \tau, \qquad \ga=\frac{4\mcH}{w^2}<0.
\eea
Due to the choice of $C_2$ one can readily check that $F_1$ is symmetric with reflection about $\tau=\frac{K(k)}{w(1-\ga)^\frac14}$, and further, so is $F_2$.  Finally, one can also see that the function $\phi(\tau,P_\phi)=\phi\left(\frac{K(k)}{2w(1-\ga)^\frac14}-\tau,-P_\phi\right)+{\rm const}$, and so the entire surface is symmetric under this interchange up to a choice of origin of $\phi$, and so can be compensated for by shifting $\phi_0$.  Note that the added reflection $P_\phi\rightarrow -P_\phi$ does not affect the functions $F_1$ or $F_2$. Essentially all we are getting is that this very symmetric situation has a reflection symmetry that relates two halves of the surface (which is why we consider this case to begin with).  Hence, if we wish to compare the connected surface to the disconnected surface, it suffices to do so for only 1 cap, and half the connected surface, which we turn to now.

\subsubsection{Gross Ooguri transition.}

First, $F_1$ at the boundary determines the radius of the Wilson loop.  This is given by
\be
F_1(\tau=0)^2=\cos^2(\theta(\tau=0))=\frac{\left(\frac{4P_\phi^2\left(P_\phi^2-\mcH\right)}{\exp\left({\frac{\sqrt{-\ga}} {(1-\ga)^{\frac14}}I_1\left(\frac{K(k)}{w(1-\ga)^\frac14}\right)}\right)} +4P_\phi^4+4P_\phi^2\mcH\right)^2-64P_\phi^6\mcH}{\left(\frac{4P_\phi^2\left(P_\phi^2-\mcH\right)}{\exp\left({\frac{\sqrt{-\ga}} {(1-\ga)^{\frac14}}I_1\left(\frac{K(k)}{w(1-\ga)^\frac14}\right)}\right)} +4P_\phi^4-4P_\phi^2\mcH\right)^2}
\ee
for the connected surface, and for the cap is given by
\be
F_1(\tau=0)^2=\cos^2(\theta_0)
\ee
and so we have solved for the single-cap solution's one parameter, $\theta_0$, that gives the same boundary conditions (at this end of the connected surface).  For the time being, we will concentrate on the case
\be
w=1
\ee
so that we are talking about a single winding, avoiding the conical singularity discussed when constructing the cap solution.\footnote{However, the conical singularity may be OK in the current context.  We are looking for the Gross Ooguri ``phase transition.''  The argument is that once the minimal surface is actually the disconnected double cap type solution, the actual interaction is dominated by a bulk field.  The site of the conical singularity may simply be an indication of where to put the vertex operator associated with this emission.}  In this case $\mcH$ and $\ga$ become redundant notation, and $\mcH=\frac{\ga}{4}$.

Next, we need to define a regulator for both surfaces.  Simply subtracting the answers is merely taking $\infty-\infty$ and so can be adjusted to be any constant: the constant depending on the regulator.  For this purpose, we introduce a radial regulator, cutting off both surfaces at $r=\frac{1}{\epsilon}$.  For the single cap solution, this happens when
\be
\tau={\rm arcsinh}\left(\frac{\epsilon}{\sqrt{1-\cos^2(\theta_0)(1+\epsilon^2)}}\right)\approx \frac{\epsilon}{{\sin(\theta_0)}}+\frac{3\cos^2(\theta_0)-1}{6\sin^3(\theta_0)}\epsilon^3+{\mathcal O}(\epsilon^5).
\ee
For the connected surface, we must consider the general expression
\be
\left(\frac{1+F_2^2}{1-F_1^2}-1\right)=\frac{1}{\epsilon^2}.
\ee
Note that if we expand the function for small values of $\tau$, we expect that $F_1$ is approximately constant, and $F_2$ goes to $\infty$ as $\frac{1}{\tau}$.  Also, we note that $I_1$ approaches zero as $\tau^3$, and so we expect $F_1^2$ to approach it's constant asymptotic as $\tau^3$.  Also, the expression for $F_1$ is an odd function, so we expect the series about $\tau=0$ to have an odd power series expansion.  Hence, expanding the left hand side of the above expression as a series in $\tau$ we should find
\be
\frac{c_{-2}^2}{\tau^2}+c_0+{\mathcal O}(\tau)=\frac{1}{\epsilon^2}.
\ee
which we may solve order by order in $\epsilon$.  For the first two terms, this is simple, and gives
\be
\tau=c_{-2}\epsilon+\frac{c_0c_{-2}}{2}\epsilon^3+{\mathcal O}(\epsilon^4).
\ee
As argued above, these first two terms come from simply expanding $F_2$ around $\tau=0$, which we may do quite simply.  Also, the asymptotic value of $F_1$ matches the $F_0=\cos(\theta_0)$ of the cap solution, so we find
\be
\tau=\frac{\epsilon}{{\sin(\theta_0)}}+\frac{3\cos^2(\theta_0)-1}{6\sin^3(\theta_0)}\epsilon^3+{\mathcal O}(\epsilon^4).
\ee
Indeed, this is because $F_2$ for the cap solution and $F_2$ for the connected solution agree for the first two orders in $\tau$ when expanding around $\tau=0$.  Next, note that the Lagrangian density will diverge like $\frac{1}{\tau^2}$, and so the action will diverge like $\frac{1}{\epsilon}$.  The above prescription for the cutoff is accurate to a large enough order that the differences between regulating with the above cutoff and the radial cutoff will go to zero as the cutoff is removed.  Hence, we use the cutoff
\be
\tau_c={\rm arcsinh}\left(\frac{\epsilon}{\sqrt{1-\cos^2(\theta_0)(1+\epsilon^2)}}\right)
\ee
which is exact for the cap solution, and close enough for the connected solution in that deviations from the exact result will vanish as $\epsilon\rightarrow 0$.  This argument suffices to show that we may do the integrals over the regulated range for $\tau$ and then take the limit of the difference.

For this, we will need to have the integrals
\bea
\frac12 A_{{\rm cap}\times 2}(\epsilon)&=&\frac{2\pi L^2}{4\pi \alpha'}\int_{\tau_c}^{\infty} d\tau \frac{1}{\sinh(\tau)^2}=\left[-\frac{\cosh(\tau)}{\sinh(\tau)}\right]_{\tau_c}^{\infty}\\
\frac12 A_{\rm connect}(\epsilon)&=&\frac{2\pi L^2}{4\pi \alpha'}\int_{\tau_c}^{\frac{K(k)}{(1-\ga)^\frac14}} d\tau \sqrt{1-\ga} \left[\frac{\dn}{\sn}\left(z,k\right)\right]^2\nn \\
&=&\frac{2\pi L^2}{4\pi \alpha'}\frac{1}{\sqrt{2k^2-1}}\Bigg[(1-k^2)z-\left[\frac{\cn\cdot \dn}{\sn}\left(z,k\right)\right] \nn \\ &&\qquad  \qquad \qquad -E(\sn(z,k),k)\Bigg]_{\tau=\tau_c}^{\tau=\frac{K(k)}{(1-\ga)^\frac14}} \\
z=(1-\ga)^{\frac14}\tau, \quad && k=\frac{\sqrt{\frac{\sqrt{1-\ga}}{2}+\frac12}}{(1-\ga)^\frac14},\quad \ga=\frac{4\mcH}{w^2}\bigg|_{w=1}=4\mcH
\eea
where we have reintroduced the radius of AdS ($L$) and the appropriate factor of $2\pi$ from $\sigma$ integration, and $E$ is the elliptic integral of the second kind.

We may then take
\bea
&& \lim_{\epsilon\rightarrow 0} \left(A_{\rm connect}(\epsilon)-A_{{\rm cap}\times 2}(\epsilon)\right) \nn \\
&&\qquad \qquad =\frac{4\pi L^2}{4\pi \alpha'}\left(1- \frac{1}{\sqrt{2k^2-1}}\left[-(1-k^2)K(k)+E(k)\right]\right)
\eea
where $E(k)$ is the complete elliptic integral of the second kind.  This goes to zero at $\mcH=0=\ga$.  This actually serves as a bit of a check: recall that when $\mcH\rightarrow 0$ the actions coincide because the two different functions $F_2$ defining the two solutions coincide.

\begin{figure}[ht!]
\centering
\includegraphics[width=0.6\textwidth]{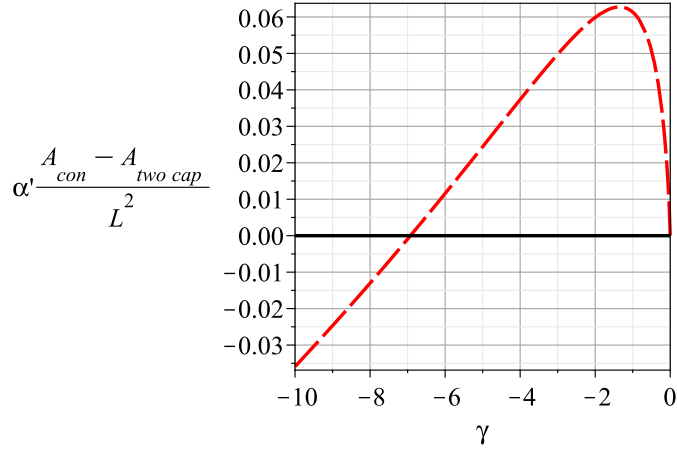}
\caption{Above we have plotted the difference of the actions for the two worldsheets (red dashed line).  We compute the critical value of $\ga$ to be $\ga_c \approx -6.92448860\rightarrow k_c=0.82317491$.  We see that for values of $\ga < \ga_c$ that the connected solution is smaller, and so preferred, and for $\ga_c<\ga<0$ that the disconnected surface is preferred.  We will later translate this into a statement about the radius of the Wilson loops and the distance between them in $\phi$.}
\label{Transition}
\end{figure}

There is in fact another zero, which we show by graphing the result in figure \ref{Transition}.
Numerically, this value is given by $\ga_c\approx -6.92448860$ which gives a critical value of $k$ being $k_c=0.82317491$.

Finally, to extract the other physical parameter defining the boundary conditions, namely $\Delta \phi$, we calculate
\bea
\tan\left(\frac{\Delta\phi}{2}\right)&=&
\tan\left(\phi\left(\tau=\frac{K(k)}{(1-\ga)^\frac14}\right)-\phi\left(\tau=0\right)\right) \nn \\
&=&\frac{\left[\exp\left({\frac{\sqrt{-\ga}} {(1-\ga)^{\frac14}}I_1\left(\frac{K(k)}{(1-\ga)^\frac14}\right)}\right)-1\right]}{\left[\exp\left({\frac{\sqrt{-\ga}} {(1-\ga)^{\frac14}}I_1\left(\frac{K(k)}{(1-\ga)^\frac14}\right)}\right)+1\right]}\frac{\sqrt{-\ga}}{2 P_\phi}
\eea
which is rather odd looking at first, given that when $P_\phi$ goes to zero, this quantity goes to infinity.  However, this is just a statement about how to get $\phi$ to go through a certain amount of angle.  In some sense, at $P_\phi=0$ the string simply ``drops'' in the $\theta$ or $r$ direction without any rotation in $\phi$.  In this situation, the string passes very close to the place where $d\phi^2$ component of the metric goes to zero, and one picks up that $\phi$ must go from $0$ to $\pi$, i.e. that half of this change is $\pi/2$, which is indeed a $\tan(\pi/2)=\infty$ limit.  Hence, somewhat counter-intuitively, $P_\phi=0$ gives the maximum $\Delta \phi$ available.  This, in fact, represents the maximally separated case for us because this is the only coordinate that separates the Wilson loops.  What we are finding is that if the two Wilson loops are separated by $\pi$, the connected minimal surface passes very close to $\theta=\frac{\pi}{2}$, given that $F_2$ does not vanish at any point.

Therefore, we have the two relations
\bea
\cos^2(\theta_0)&=&\frac{\left(\frac{P_\phi^2\left(4P_\phi^2-\ga\right)}{\exp\left({\frac{\sqrt{-\ga}} {(1-\ga)^{\frac14}}I_1\left(\frac{K(k)}{(1-\ga)^\frac14}\right)}\right)} +4P_\phi^4+P_\phi^2\ga\right)^2-16P_\phi^6\ga}{\left(\frac{P_\phi^2\left(4P_\phi^2-\ga\right)}{\exp\left({\frac{\sqrt{-\ga}} {(1-\ga)^{\frac14}}I_1\left(\frac{K(k)}{(1-\ga)^\frac14}\right)}\right)} +4P_\phi^4-P_\phi^2\ga\right)^2} \\
\tan\left(\frac{\Delta\phi}{2}\right)&=&\frac{\left[\exp\left({\frac{\sqrt{-\ga}} {(1-\ga)^{\frac14}}I_1\left(\frac{K(k)}{(1-\ga)^\frac14}\right)}\right)-1\right]}{\left[\exp\left({\frac{\sqrt{-\ga}} {(1-\ga)^{\frac14}}I_1\left(\frac{K(k)}{(1-\ga)^\frac14}\right)}\right)+1\right]}\frac{\sqrt{-\ga}}{2 P_\phi}
\eea
where on the left we have natural objects to define the boundary conditions $F_0$ and $\Delta \phi$ and on the right, we have the two constants associated with the integrals of motion $\ga$ and $P_\phi$ (and recall that $k$ is a function of $\ga$).  Now, we solve these equations and find
\bea
\frac{2P_\phi}{\sqrt{-\ga}}&=&\frac{\cos(\theta_0)}{\sqrt{\sin^2(\theta_0)+\tan^2\left(\frac{\Delta\phi}{2}\right)}} \label{solveRat}\\
&& \kern -5 em \exp\left({\frac{2k\sqrt{1-k^2}}{\sqrt{2k^2-1}}I_1\left(\sqrt{2k^2-1}K(k)\right)}\right) \nn \\ &=&\frac{\sqrt{\sin^2(\theta_0)+\tan^2\left(\frac{\Delta\phi}{2}\right)}+\tan\left(\frac{\Delta\phi}{2}\right)\cos(\theta_0)} {\sqrt{\sin^2(\theta_0)+\tan^2\left(\frac{\Delta\phi}{2}\right)}-\tan\left(\frac{\Delta\phi}{2}\right)\cos(\theta_0)}  \label{solveI1}\\
I_1\left(\sqrt{2k^2-1}K(k)\right)&=&K(k)-\frac{1-k^2}{k^2}\Pi\left(\frac{2k^2-1}{k^2},k\right) \label{I1Bndy}
\eea
where we have assumed a positive $P_\phi$, which gives positive $\tan\left(\frac{\Delta \phi}{2}\right)$.
Note that the quantity on the right is strictly larger than zero, and larger than one.  This serves as a check because $I_1$ was defined to be positive definite for positive definite argument, and so the left hand side is strictly greater than one as well.  To get a negative $P_\phi$, one needs to simply reverse the sign of the right hand side of (\ref{solveRat}) because $\cos(\theta_0)>0$, and further take the reciprocal of the right hand side of (\ref{solveI1}).  Then, however, $\Delta\phi$ changes sign, so again, the new right hand side of (\ref{solveI1}) is bigger than one, again agreeing with the behavior of $I_1$.  This can be accounted for by simply taking the expression (\ref{solveI1}) and replacing $\tan\left(\frac{\Delta \phi}{2}\right)\rightarrow \left|\tan\left(\frac{\Delta \phi}{2}\right)\right|$, resulting in the expression (\ref{introktobndydata}) given in the introduction.
In fact, the left hand side of (\ref{solveI1}) monotonically increasing for increasing $\ga$ and the right hand side is monotonically increasing for increasing $\left|\tan\left(\frac{\Delta \phi}{2}\right)\right|$ and fixed $\theta_0$ (in fact, this expression is completely invertible).  This gives that given $0<\theta_0<\frac{\pi}{2}$ that increasing $\ga$ (decreasing it's magnitude, because $\ga<0$) increases the $\Delta \phi$ between the two Wilson loops.  This gives the qualitatively right behavior.  When the loops are close ($\ga\rightarrow -\infty$) the connected surface is preferred, and when they are taken further apart ($\ga\rightarrow 0$) the disconnected surface is preferred.  This happens at
\bea
&& \exp\left({\frac{2k\sqrt{1-k^2}}{\sqrt{2k^2-1}}I_1\left(\sqrt{2k^2-1}K(k)\right)}\right)\Bigg|_{k_c}\equiv {\mathcal E} \approx 2.40342513\nn\\
 &&=\frac{\sqrt{\sin^2(\theta_0)+\tan^2\left(\frac{\Delta\phi}{2}\right)}+\tan\left(\frac{\Delta\phi}{2}\right)\cos(\theta_0)} {\sqrt{\sin^2(\theta_0)+\tan^2\left(\frac{\Delta\phi}{2}\right)}-\tan\left(\frac{\Delta\phi}{2}\right)\cos(\theta_0)}
\eea
or solving
\be
\tan^2\left(\frac{\Delta\phi_c}{2}\right)=\frac{\sin^2(\theta_0)}{\frac{({\mathcal E}+1)^2}{({\mathcal E}-1)^2}\cos^2(\theta_0)-1} \label{deltaphicrit}
\ee
determines the critical separation distance where the disconnected surface is preferred.  One may take a flat space limit by considering small loops $\theta\rightarrow 0$.  This gives
\be
(\Delta \phi_c)^2 = \frac{({\mathcal E}-1)^2}{\mathcal E} \theta^2.
\ee
In this approximation, one can view the left hand side $\Delta \phi$ as being the distance of a Cartesian coordinate, and the right hand side $\theta$ as being the radius of the Wilson loops.  Both sides have had units removed with the $AdS$ radius.  Hence, we interpret the above as saying
\be
(\Delta x_c)^2 = \frac{({\mathcal E}-1)^2}{\mathcal E} R^2
\ee
where $R$ is the radius of both Wilson loops, and $\Delta x$ is the distance between their centers.  Numerically, this gives $\Delta x_c = 0.90524981 R$, agreeing with the flat space result quoted in \cite{Zarembo:1999bu}.

There is one more interesting feature of the above equations.  Given that $\mathcal E$ is a constant, one may wonder whether there are sufficiently large Wilson loops such that the critical distance $\Delta \phi$ becomes maximal.  Indeed, this happens when
\bea
(1-F_{0,{\rm crit}}^2)=\sin^2(\theta_{\rm crit})&=&1-\frac{({\mathcal E}-1)^2}{({\mathcal E}+1)^2}\approx 0.8299619582 \nn \\
\theta_{\rm crit}&\approx& 0.36470575 \pi
\eea
and so these loops are actually quite large, becoming an appreciable fraction of the maximal angle $\theta=\frac12 \pi$. We think of this effect as being similar to Hawking-Page \cite{Hawking:1982dh}, in that the scale associated with the sphere appears to play a crucial role.


\section{Discussion}
\label{discussion}

Here we discuss possible future directions.
\begin{enumerate}
\item Note that we may describe ``orientation reversed'' Wilson loop correlators, given the structure of the solutions for $\mcH>0$.  In these cases the string world-volume reflects through an origin, and reverses the orientation: this reverses the orientation of the second Wilson loop, and so computes a correlator of the form $\langle W W^\dagger\rangle$.  This reverse orientation may have an interesting effect on which minimal surface is preferred, and so affect the Gross-Ooguri transition.
\item It would be interesting to calculate the asymptotic behavior for small Wilson loops largely separated using the operator product expansion approach of \cite{Berenstein:1998ij} adapted to S$^3$.  This answer should agree with the flat space result \cite{Berenstein:1998ij} for a certain regime, but should differ once the separation distance becomes comparable to the radius of the sphere.
\item It would also be interesting to see how the transition \cite{Olesen:2000ji} occurs for unequal size Wilson loops on S$^3$, and see how this affects the ``critical size'' calculation done here for equal size Wilson loops.
\item Having a profile on the S$^5$ should not change the behavior of the surfaces in AdS drastically: one can see this by noting the effect of treating $H_1$ and $H_2$ as both being non zero only affects certain coefficients in the equations of motion, rather than changing their structure.  One expects the sphere equations of motion to completely decouple except for this effect (see e.g. \cite{Dorn:2009hs} for a similar story for the gluon scattering minimal surfaces).
\item One could also calculate correlators from insertion of other types of operators in the spirit of \cite{Berenstein:1998ij,Zarembo:2002ph,Pestun:2002mr,Tsuji:2006zn,Miwa:2006vd,Sakaguchi:2007ea}, and try to extract information about vertex operator correlators, as in \cite{Buchbinder:2010vw,Roiban:2010fe,Ryang:2010bn,Hernandez:2010tg}.
\item These solutions could also be used as a leading order ``vaccuum'' for soliton transformations.  It would also be interesting to try to ``add'' multiple solutions together using generalized superposition principles sometimes available in integrable systems.
\item It would also be interesting to see how finite size effects of the S$^3$ affect the behavior of small vs. large Wilson loops in the weak coupling approximation.

\end{enumerate}

\section*{Acknowledgements}

We wish to thank C\'esar A. Terrero Escalante who was involved in early stages of this work.  We are grateful to Radu Roiban for general comments.  BB is grateful to Erich Poppitz for many illuminating discussions, and to Nadav Drukker for comments.  This work was supported in part under a grant from NSERC of Canada, and supported in part by the US Department of Energy under grant DE-FG02-95ER40899.

\appendix

\section{Finding $F_1$.}

\label{OtherConstantsF1}

\subsection{general considerations}

Now that we have $F_2$, we may turn to finding $F_1$.  First, we take an $F_2$ derivative of $F_1$ by dividing the $F_1$ equation by the $F_2$ equation to shed light on the final form of the solution:
\be
\frac{dF_1}{dF_2}=\frac{1-F_1^2}{1+F_2^2}\frac{\sqrt{-\mcH+P_\phi^2+P_t^2-P_t^2F_1^2-P_\phi^2\frac{1}{F_1^2}}}{ \sqrt{\mcH+w^2F_2^2+w^2+F_2^4}}.
\ee
Therefore we need to perform two integrals
\bea
&&\int \frac{dF_1}{(1-F_1^2)\sqrt{-\mcH+P_\phi^2+P_t^2-P_t^2F_1^2-P_\phi^2\frac{1}{F_1^2}}} \nn \\
&&\qquad \qquad  = \int \frac{dF_2}{(1+F_2^2)\sqrt{\mcH+w^2F_2^2+w^2+F_2^4}}=\int\frac{d\tau}{(1+F_2^2)}.
\eea
The $F_1$ integral is what we would need to perform whether or not we had divided the equations, and the $F_2$ integral tells us the form of the final answer: it is some elliptic pi function (\ref{defPi}).
It is clear that the $F_2$ integral must be real when it is represented as a $\tau$ integral, and so the integration for the $F_1$ function must also yield a real result if $F_1$ is to be real.

Therefore, let us check under what conditions one can find a real $F_1$.  It is quite clear that the $F_1$ integral is only real if
\be
P_t^2+P_\phi^2-\mcH>0.
\ee
However, we can be more stringent simply by exploring the term under the square root more closely:
\be
-P_t^2F_1^4+(P_t^2+P_\phi^2-K)F_1^2-P_{\phi}^2
\ee
where we have multiplied by an $F_1^2$ for convenience (this has no effect on the sign of the above if $F_1$ is real).
We must have some sections where it is possible for $F_1$ to be real: this happens where the above quantity is bigger than zero.  This is easy to check simply by solving the above equation as a quadratic equation in $F_1^2$.  Solving this we find
\bea
&&\frac{P_t^2+P_\phi^2-\mcH - \sqrt{(P_t^2+P_\phi^2-\mcH)^2-4P_\phi^2P_t^2}}{2P_t^2}\leq  F_1^2 \leq \nn \\
&&\qquad \qquad \frac{P_t^2+P_\phi^2-\mcH + \sqrt{(P_t^2+P_\phi^2-\mcH)^2-4P_\phi^2P_t^2}}{2P_t^2} \\
&&\rightarrow \frac{P_t^2+P_\phi^2-\mcH - \sqrt{\left((P_t-P_\phi)^2-\mcH\right)\left((P_t+P_\phi)^2-\mcH\right)}}{2P_t^2}\leq F_1^2 \leq \nn \\
&& \qquad \qquad \frac{P_t^2+P_\phi^2-\mcH + \sqrt{\left((P_t-P_\phi)^2-\mcH\right)\left((P_t+P_\phi)^2-\mcH\right)}}{2P_t^2}.\label{quadSolve1}
\eea
One of these solutions must be real and positive for there to be a region where $F_1$ can be real.  In fact, because we know that $P_t^2+P_\phi^2-\mcH>0$ all that needs to be shown (by the first line above) is that
\be
(P_t^2+P_\phi^2-\mcH)^2-4P_\phi^2P_t^2>0.
\ee
If this is the case, it is obvious that the $+$ square root gives a positive value, but also the $-$ value does also because $\sqrt{(P_t^2+P_\phi^2-\mcH)^2-4P_\phi^2P_t^2}$ is smaller in magnitude than $(P_t^2+P_\phi^2-\mcH)$.  The above inequality becomes
\be
|P_t^2+P_\phi^2-\mcH|>2|P_t| |P_\phi|, \label{absineq1}
\ee
however we have already seen that $P_t^2+P_\phi^2-\mcH>0$ so the absolute value on the left hand side does nothing.  This gives, finally, that
\be
(|P_t|-|P_\phi|)^2>\mcH.
\ee
This restriction is more stringent than $P_t^2+P_\phi^2-\mcH>0$, as seen from (\ref{absineq1}), and so the above restriction replaces it.  The above restriction can, in fact, be read from the second line of (\ref{quadSolve1}) simply by considering the two cases where $P_t P_\phi$ is either positive or negative.  In either case, one of the factors under the square root is positive, and so the sign under the square root is determined by the other factor: precisely the one that reads $(|P_t|-|P_\phi|)^2- \mcH$.  We will see in a moment that these both are eliminated by an even more stringent constraint.

There is one more important note to make: the function $-1<F_1=\frac{r\cos(\theta)}{1+r^2}<1$.  Above, we have just noted that for $F_1$ to be real we must have $F_1^2$ between the two solutions (\ref{quadSolve1}).  The smaller of the two of these must be less than one to get an $F_1$ that comes from a real $r$ and real $\theta$.  Hence we find that
\bea
&& \frac{P_t^2+P_\phi^2-\mcH - \sqrt{(P_t^2+P_\phi^2-\mcH)^2-4P_\phi^2P_t^2}}{2P_t^2}<1 \nn \\
\rightarrow && \frac{P_\phi^2-P_t^2-\mcH - \sqrt{(P_\phi^2-P_t^2-\mcH)^2-4P_t^2\mcH}}{2P_t^2}<0 \label{radineq1}
\eea
We have already seen that the square root defined above must be real, via the constraint $(|P_t|-|P_\phi|)^2> \mcH$.

The above constraint breaks into two cases depending on the sign of $\mcH$.  When $\mcH>0$ the first term dominates in magnitude compared to the term under the square root.  This means that the first term determines the sign, so we find that $P_t^2-P_\phi^2>-\mcH$.  Combining this with $(|P_t|-|P_\phi|)^2> \mcH$ we find
\be
|P_t|-|P_\phi|>0\qquad \mbox{for $\mcH>0$}.
\ee
This, therefore, tells us which square root to take from the previous inequality and we find
\be
|P_t|-|P_\phi|>\sqrt{\mcH} \qquad \mbox{for $\mcH>0$}
\ee
where above we take the $+$ square root only.  The $\mcH<0$ is trivial because the square root term in (\ref{radineq1}) dominates, and so the inequality is automatically satisfied.

So, again, the above constraint is more stringent than all other constraints.  Further note that the above is a strict inequality: if this number is equal to zero, then there is no window of allowed values for $F_1$: this would give that $F_2=1$ strictly.  This is only possible for $r\rightarrow \infty$ and so does not describe a true solution.  In fact, we can see the inequality in the $\mcH=0$ limit:
\be
\frac{P_\phi^2-P_t^2 - \sqrt{(P_\phi^2-P_t^2)^2}}{2P_t^2}<0.
\ee
This has two distinct cases: if $P_\phi^2>P_t^2$ then the above is equal to 0, and we get no solutions;  if $P_\phi^2<P_t^2$ then we see that the above is $-\frac{(P_t^2-P_{\phi}^2)}{P_t^2}<0$ which is consistent with the assumption.  One can see this constraint from the exact solution given by the equation for $\theta$ in (\ref{K0th}).  This also excludes the possibility that $|P_\phi|=|P_t|, \mcH=0$: one can check this easily by plugging directly into the integrand finding $\frac{dF_1}{(1-F_1^2)\sqrt{\frac{-P_t^2(1-F^2)^2}{F^2}}}$, which is manifestly imaginary.

Finally, performing explicitly the $F_1$ integral, one finds
\bea
&&\int\frac{dF_1}{(1-F_1^2)\sqrt{-\mcH+P_\phi^2+P_t^2-P_t^2F_1^2-P_\phi^2\frac{1}{F_1}^2}} \nn \\
&&\qquad \qquad \qquad \qquad =-\frac{{\rm arctanh }\left(\frac{2\sqrt{-\mcH}\sqrt{-P_t^2F_1^4+(P_t^2+P_\phi^2-K)F_1^2-P_{\phi}^2}} {P_{\phi}^2-P_t^2+\mcH+(P_t^2-P_{\phi}^2+K)F_1^2}\right)}{2\sqrt{-{\mcH}}} \label{F1int}
\eea
up to a constant (that we leave in the $F_2$ integral).  Consider first $\mcH<0$.  Then, one can see that comparing the magnitudes of the numerator and denominator of the argument of ${\rm arctanh}$
\bea
&& \left(P_{\phi}^2-P_t^2+\mcH+(P_t^2-P_{\phi}^2+K)F_1^2\right)^2 \nn \\
&&\qquad -4(-\mcH)\left(-P_t^2F_1^4+(P_t^2+P_\phi^2-K)F_1^2-P_{\phi}^2\right) \nn \\
&& \qquad =\left((P_t^2+P_\phi^2-\mcH)^2-4P_t^2P_\phi^2\right)(1-F_2)^2>0.
\eea
The result of the ${\rm arctanh}$ is therefore real because the argument lies in the interval $(-1,1)$.

When $\mcH>0$, we see that we must push some factors of $i$ through, which changes the function to an $\arctan$.  An $\arctan$ can accept any real input and so, other than a reality condition (which we have already met), the above is well defined.  However, the output lies in a defined range ($\left(-\frac{\pi}{2},\frac{\pi}{2}\right)$, and one may think that this constrains the value of the integration constant in $F_2$.  We will address this when the time comes, and find that it is not an issue.

There is an extremely important caveat for the above arguments to hold: that we have not divided by a function that is identically 0.  This case deserves special consideration.  We can immediately see that if this term is zero, we have a constant $F_1$.  In such a case, we find that the
\be
F_1^2=\frac{P_t^2+P_\phi^2-\mcH\pm\sqrt{(P_t^2+P_\phi^2-\mcH)^2-4P_\phi^2P_t^2}}{2P_t^2}.
\ee
For this, we have that one of these solutions must satisfy $F_1^2<1$ as well, so that the above restrictions found must also hold.  However, we must also explore the $+$ square root, and see when it is possible for this to be less than one. We therefore explore
\bea
0<\frac{P_t^2+P_\phi^2-\mcH + \sqrt{(P_t^2+P_\phi^2-\mcH)^2-4P_\phi^2P_t^2}}{2P_t^2}<1 \nn \\
-1<\frac{-P_t^2+P_\phi^2-\mcH + \sqrt{(-P_t^2+P_\phi^2-\mcH)^2-4P_t^2\mcH}}{2P_t^2}<0.
\eea
When $\mcH<0$, the bottom line is clearly violated on the upper bound (given a real square root).  When $\mcH>0$, the top lines lower bound is obviously satisfied, and the lower lines upper bound is satisfied if $P_t^2-P_\phi^2+\mcH>0$: this condition is the same as when the square root is the other sign, and so $|P_t|-|P_\phi|> \sqrt{\mcH}$.

To conclude, we summarize the restrictions found in this section
\bea
\mbox{For $F_1'\neq 0$:}&& \qquad  \mcH<0: \qquad \mbox{No restrictions} \nn \\
&& \qquad  \mcH \geq 0: \qquad |P_t|-|P_\phi|>\sqrt{\mcH} \nn \\
\mbox{For $F_1'=0$:}  && \qquad F_1^2=\frac{P_t^2+P_\phi^2-\mcH\pm\sqrt{(P_t^2+P_\phi^2-\mcH)^2-4P_\phi^2P_t^2}}{2P_t^2} \nn \\
&& \mcH <0 : \mbox{minus sign only, no other restrictions} \nn \\
&& \mcH >0 : |P_t|-|P_\phi|>\sqrt{\mcH}\quad  \mbox{either sign OK}
\eea
Note that the other possible solution $F_1^2=\frac{r^2\cos^2(\theta)}{1+r^2}=1$ for $F_1'=0$ gives that the string is exactly on the boundary, and so this is not an interesting solution.

\subsection{Finding $F_1$ for $\mcH<0$, $\gamma<0$}

Next, let us name a function $g(\tau)$, given in $\ref{F1int}$
\bea
&&\left(\frac{2\sqrt{-\mcH}\sqrt{-P_t^2F_1^4+(P_t^2+P_\phi^2-\mcH)F_1^2-P_{\phi}^2}} {P_{\phi}^2-P_t^2+\mcH+(P_t^2-P_{\phi}^2+K)F_1^2}\right)\nn \\
&&=g(\tau)\equiv-\tanh\left(2\sqrt{-\mcH}\int\frac{d\tau}{1+F_2^2}\right)
\eea
where the $F_2$ integral is solved the same way as in the text.  One may solve the above for $F_1$ to find
\bea
F_1^2&=&\Bigg[\left((P_t^2-P_\phi^2)^2-\mcH^2\right)g(\tau)^2-2\mcH(P_t^2+P_\phi^2-\mcH) \nn \\
&&\quad \pm 2\mcH\sqrt{(1-g(\tau)^2)\left((P_t^2+P_\phi^2-\mcH)^2-4P_t^2P_\phi^2\right)}\Bigg] \nn \\
&&\qquad \div \left[(P_t^2-P_\phi^2+\mcH)^2g(\tau)^2-4P_t^2\mcH\right]
\eea
One may check that the expression in strictly greater than zero because the term under the square root has smaller magnitude than the term without a square root (the difference of squares is positive) for the allowed values of $0<g(\tau)^2<1$.  Next, one may compute
\bea
F_1^2-1&=&\Bigg[-2\mcH(P_t^2-P_\phi^2+\mcH)g(\tau)^2-2\mcH(P_\phi^2-P_t^2-\mcH) \nn \\
&& \quad \pm 2\mcH\sqrt{(1-g(\tau)^2)\left((P_t^2+P_\phi^2-\mcH)^2-4P_t^2P_\phi^2\right)}\Bigg] \nn \\
&&\qquad \div \left[(P_t^2-P_\phi^2+\mcH)^2g(\tau)^2-4P_t^2\mcH\right].
\eea
In this, one can show that the term with the square root dominates, and so this term completely determines the sign.  Therefore, we must take the $+$ sign above (because $\mcH<0$) to have $F_1^2<1$.  Hence, we find
\bea
F_1^2&=&\Bigg[\left((P_t^2-P_\phi^2)^2-\mcH^2\right)g(\tau)^2-2\mcH(P_t^2+P_\phi^2-\mcH) \nn \\
&&\quad + 2\mcH\sqrt{(1-g(\tau)^2)\left((P_t^2+P_\phi^2-\mcH)^2-4P_t^2P_\phi^2\right)}\Bigg] \nn \\
&&\qquad \div \left[(P_t^2-P_\phi^2+\mcH)^2g(\tau)^2-4P_t^2\mcH\right]
\eea
where this expression is manifestly positive and less than 1 for $0<g(\tau)^2<1$.

\section{Elliptic Functions}

\label{ellipticSection}

We define the elliptic integrals
\bea
F(T,k)\equiv \int_0^{T}dt\frac{1}{\sqrt{1-k^2t^2}\sqrt{1-t^2}} \label{defF}\\
E(T,k)\equiv \int_0^{T}dt\frac{\sqrt{1-k^2t^2}}{\sqrt{1-t^2}} \label{defE} \\
\Pi(T,\alpha^2,k)\equiv \int_0^T \frac{dt}{(1-\alpha^2t^2)\sqrt{(1-t^2)(1-k^2t^2)}} \label{defPi}
\eea
and the complete elliptic integrals as
\bea
K(k)\equiv F(1,k) \\
E(k)\equiv E(1,k) \\
\Pi(\alpha^2,k)=\Pi(1,\alpha^2,k).
\eea
For the Jacobi elliptic functions, we follow the conventions of \cite{Whit,Grad,NIST}.  The reference \cite{Whit} has several forms of the above integrals useful for defining various combinations of the Jacobi elliptic functions, while \cite{Grad} has extremely useful tables for their transformation properties under shifts, and \cite{NIST} has some useful transformation properties of $\Pi(T,\alpha^2,k)$ not contained in the other references.

\end{document}